\documentclass[prc,twocolumn,showpacs,preprintnumbers]{revtex4}
\usepackage{bm}
\usepackage{amssymb}
\usepackage{amsmath}
\usepackage{graphicx}

\begin{document}
\title{The structure of accreted neutron star crust}
\author{C. J. Horowitz}\email{horowit@indiana.edu} 
\affiliation{Department of Physics and Nuclear Theory Center,
             Indiana University, Bloomington, IN 47405}
\author{D. K. Berry}\email{dkberry@indiana.edu}
\affiliation{University Information Technology Services,
             Indiana University, Bloomington, IN 47408}

\date{\today}
\begin{abstract}
Using molecular dynamics simulations, we determine the structure of neutron star crust made of rapid proton capture nucleosynthesis material.  We find a regular body centered cubic lattice, even with the large number of impurities that are present.   Low charge $Z$ impurities tend to occupy interstitial positions, while high $Z$ impurities tend to occupy substitutional lattice sites.  We find strong attractive correlations between low $Z$ impurities that could significantly increase the rate of pycnonuclear (density driven) nuclear reactions.  The thermal conductivity is significantly reduced by electron impurity scattering.  Our results will be used in future work to study the effects of impurities on mechanical properties such as the shear modulus and breaking strain.
\end{abstract}
\smallskip
\maketitle

\section{Introduction}
Neutron stars, collapsed objects half again more massive than the sun, are thought to have solid crusts about a kilometer thick.  Detailed properties of this crust are important for many X-ray, radio, and gravitational wave observations.  Because of the great densities, the electronic structure of the crust is likely very simple, consisting of an extremely degenerate relativistic Fermi gas.  The system can be modeled as nearly classical ions interacting via screened Coulomb, or Yukawa, interactions, where the screening length $\lambda$ depends on the electron density.

Indeed many condensed matter systems can be modeled with Yukawa interactions and much is known about the properties of a single component Yukawa system.  See for example ref. \cite{yukawasystems}.  Because the ion-ion interaction is purely repulsive, there is no liquid-gas phase transition.  However there is a liquid solid phase transition at a melting temperature $T_m$ that depends primarily on the Coulomb parameter $\Gamma$,
\begin{equation}
\Gamma=\frac{Z^2e^2}{a T}\, .
\label{Gamma}
\end{equation}   
Here the ions have charge $Z$, $T$ is the temperature and the ion sphere radius $a$ is
\begin{equation}
a=\Bigl[\frac{3}{4\pi n}\Bigr]^{1/3}
\label{a}
\end{equation}
with $n$ the ion density.The system melts at a temperature for which $\Gamma\approx 175$ (assuming the screening length is relatively large).

Neutron stars, that accrete material from a binary companion, form new crust from the ashes of nuclear reactions.  Simulations of rapid proton capture nucleosynthesis \cite{rp1}\cite{rp2} find a complex composition with many different ion species.  These species then undergo electron capture as the material is buried by further accretion to greater densities \cite{gupta}.  This still leaves a complex composition of very neutron rich isotopes with many different chemical elements.  In this paper, we investigate the structure of the resulting solid crust when this complex mixture freezes. 

Monte Carlo simulations \cite{mcocp} of the freezing of a classical one component plasma (OCP) indicate that it can freeze into imperfect body centered cubic (bcc) or face-centered cubic (fcc) microcrystals.  Unfortunetly not much has been published on the freezing of a multi-component plasma (MCP), although Wunsch et al. study the structure of a MCP liquid \cite{MCPliquid}.  There are many possibilities for the state of a cold MCP \cite{yakovlev}.  It can be a regular MCP lattice; or microcrystals; or an amorphous, uniformly mixed structure; or a lattice of one phase with random admixture of other ions; or even an ensemble of phase separated domains.

One possibility is that impurities could become frozen into random configurations that only relax on very long time scales.  This could lead to the formation of a glass.  For example, the binary Lennard-Jones system, with two species of different sizes, can form a glass \cite{LJglass}.  Here the hard core of the interaction keeps the different species from diffusing.   However the screened coulomb interaction has a relatively soft $1/r$ core.  This may allow impurities to diffuse and prevent the formation of a glass.  

In this paper we use molecular dynamics simulations to calculate radial distribution functions $g(r)$ and static structure factors $S(q)$ to determine the structure of the crust.  In previous work we determined the chemical separation that takes place as the crust freezes.  We found that the liquid phase is greatly enriched in low charge $Z$ ions while the solid is enriched in high $Z$ ions \cite{horowitz}.  The distribution of low $Z$ (impurity) ions in the crust may be important for the rate of strongly screened thermonuclear or pycnonuclear (density driven) reactions.  In ref. \cite{pycno} we found that fusion of $^{24}$O + $^{24}$O could be an important heat source in the crust.  

In ref. \cite{thermalcond} we calculated the thermal conductivity of the crust.  We found the crust to be a regular crystal with a relatively high thermal conductivity.  Recently the cooling of two neutron stars has been observed after extended outbursts \cite{Wijnands, cackett}.  These outbursts heat the stars' crusts out of equilibrium and then the cooling time is measured as the crusts return to equilibrium.  The surface temperature of the neutron star in KS 1731-260 decreased with an exponential time scale of 325 $\pm$ 100 days while MXB 1659-29 has a time scale of 505 $\pm$ 59 days \cite{cackett}.  Comparing these observations, of rapid cooling, to calculations by Rutledge et al. \cite{rutledge}, Shternin et al. \cite{shternin}, and Brown et al. \cite{cumming08}  strongly suggest that the crust has a high thermal conductivity.  

The shear modulus of a single component system was calculated in ref. \cite{shearmod} where electron screening was found to reduce the shear modulus by about 10\% compared to a pure $1/r$ Coulomb system \cite{ogata}.  The shear modulus determines the frequency of shear oscillations of the neutron star crust.  These may have been observed as quasiperiodic oscillations in magnetar giant flares \cite{qpos}.  In the future, we will use the results of this paper to calculate the effects of impurities on the shear modulus.

The breaking strain is the deformation of the crust when it fails.  This determines the maximum height of mountains on the surface of neutron stars.  These may be important sources of gravitational waves for rapidly rotating stars \cite{gw}.  The breaking strain may also be important for star quake models of magnetar giant flares \cite{thompsonduncan}.  In ref. \cite{kadau} we examine the effects of impurities on the breaking strain.

The present paper is similar to ref. \cite{thermalcond}, however here we use a significantly different composition.  Ref. \cite{thermalcond} had so many low $Z$ impurities that phase separation was apparently taking place.  Therefore, ref. \cite{thermalcond} results may not be directly applicable to large uniform systems.  In the present paper we consider a system with fewer impurities, as described in Section \ref{MD}, that may form a uniform system.

Impurities can limit the thermal conductivity.  If the impurities are weakly correlated then their effect on the thermal conductivity can be described by an impurity parameter $Q$ \cite{impurities},
\begin{equation}
Q=(\Delta Z)^2=\ \langle Z^2 \rangle - \langle Z \rangle^2.
\label{Q}
\end{equation}
This depends on the dispersion in the charge $Z$ of each ion.  The rp process ash composition of ref. \cite{gupta} and ref. \cite{horowitz} has a relatively large value of $Q=38.9$.  In this paper, the composition we use has a smaller, but still significant, value $Q=22.54$.  This lower value is because of a reduction in low $Z$ impurities.  Impurity scattering can be important at low temperatures where there is less scattering from thermal fluctuations.  Note that ref. \cite{impurities} assumes the impurities are weakly correlated.  If there are important correlations among the impurities, for example if there is a tendency for low $Z$ ions to cluster together instead of being distributed at random through out the lattice, then the effects of impurities on the thermal conductivity could be different from what is calculated in ref. \cite{impurities}.  In this paper we perform MD simulations to study the distribution of impurities and their effect on the conductivity.

In section \ref{MD} we describe our molecular dynamics simulations.   Results for the radial distribution function $g(r)$, the static structure factor $S(q)$, and the thermal conductivity are presented in section \ref{Results}.  We conclude in section \ref{Conclusions}.

\section{Molecular Dynamics Simulations}
\label{MD}
In this section we describe our classical molecular dynamics simulations.  We begin with a discussion of the composition.  
Schatz et al. have calculated the rapid proton capture (rp) process of hydrogen burning on the surface of an accreting neutron star \cite{rp1}, see also \cite{rp2}.  This produces a variety of nuclei up to mass $A\approx 100$.  Gupta et al. then calculate how the composition of this rp process ash evolves, because of electron capture and light particle reactions, as the material is buried by further accretion.  Their final composition, at a density of $2.16\times 10^{11}$ g/cm$^3$, has forty \% of the ions with atomic number $Z=34$, while an additional 10\% have $Z=33$.  The remaining 50\% have a range of lower $Z$ from 8 to 32.  In particular about 3\% is $^{24}$O and 1\% $^{28}$Ne.  This Gupta et al. composition is listed in the mixture column of Table I in ref. \cite{horowitz} and was used for the simulations in ref. \cite{thermalcond}.  In general, nuclei at this depth in the crust are expected to be neutron rich because of electron capture.

Material accretes into a liquid ocean.  As the density increases near the bottom of the ocean, the material freezes.  However we found chemical separation when the complex rp ash mixture freezes \cite{horowitz}.  The ocean is greatly enriched in low $Z$ elements compared to the newly formed solid.  What does chemical separation mean for the structure of the crust?  Here we make a very simple assumption and use the composition of the solid phase that was found in ref. \cite{horowitz}.  [Note that for simplicity we drop chemical elements with number fraction less than 0.001.]  This composition is listed in Table \ref{tableone} and is depleted in low $Z$ elements compared to the original Gupta et al. composition.   For example, we now have only about 1\% $^{24}$O compared to the original 3\%.   Our composition may not be self consistent because we expect chemical separation to enrich the ocean in low $Z$ elements and this may change the composition of the newly formed crust.  This should be investigated in future work.

\begin{table}
\caption{Composition of rapid proton capture nucleosynthesis ash MD simulations: number fraction $x_i$ of chemical element with atomic number $Z$ and mass number $A$.} 
\begin{tabular}{lll}
$Z$ & $A$ & $x_i$ \\
\toprule
8 & 24 & 0.0093 \\
10 & 28 & 0.0023 \\
20 & 62 & 0.0023 \\
22 & 66 & 0.0625 \\
24 & 74 & 0.0625 \\
26 & 76 & 0.1019 \\
27 & 77 & 0.0023 \\
28 & 80 & 0.0741 \\
30 & 90 & 0.0949 \\
32  & 96 & 0.0139 \\
33 & 99 & 0.1389 \\
34 & 102 & 0.4306 \\
36 & 106 & 0.0023 \\
47 & 109 & 0.0023 \\
\end{tabular} 
\label{tableone}
\end{table}

The electrons form a very degenerate relativistic electron gas that slightly screens the interaction between ions.  We assume the potential $v_{ij}(r)$ between the ith and jth ion is,
\begin{equation}
v_{ij}(r) = \frac{Z_i Z_j e^2}{ r} {\rm e}^{-r/\lambda_e}\, ,
\label{vij}
\end{equation}
where $r$ is the distance between ions and the electron screening length is $\lambda_e=\pi^{1/2}/[2e(3\pi^2 n_e)^{1/3}]$.  Here $n_e$ is the electron density.  Note that we do not expect our results to be very sensitive to the electron screening length.  For example, the OCP melting point that we found in ref. \cite{horowitz}, using a finite $\lambda_e$, agrees well with the result for $\lambda_e=\infty$. 

To characterize our simulations , we define an average Coulomb coupling parameter $\Gamma$ for the MCP,
\begin{equation}
\Gamma= \frac{\langle Z^{5/3} \rangle \langle Z \rangle^{1/3} e^2}{a T}\, ,
\label{gamma}
\end{equation}
where the ion sphere radius is $a=(3/4\pi n)^{1/3}$ and $n=n_e/\langle Z\rangle$ is the ion density.   The OCP freezes near $\Gamma=175$.  In ref. \cite{horowitz} we found that the impurities in our MCP lowered the melting temperature until $\Gamma=247$.  Finally, we can measure time in our simulation in units of one over an average plasma frequency $\omega_p$,
\begin{equation}
\omega_p=\Bigl(\sum_j\frac{Z_j^2 4\pi e^2 x_j n}{M_j}\Bigr)^{1/2}\, ,
\label{omegap}
\end{equation}
where $M_j$ is the average mass of ions with charge $Z_j$ and abundance $x_j$ (by number).

\begin{figure}[ht]
\begin{center}
\includegraphics[width=3in,angle=0,clip=true] {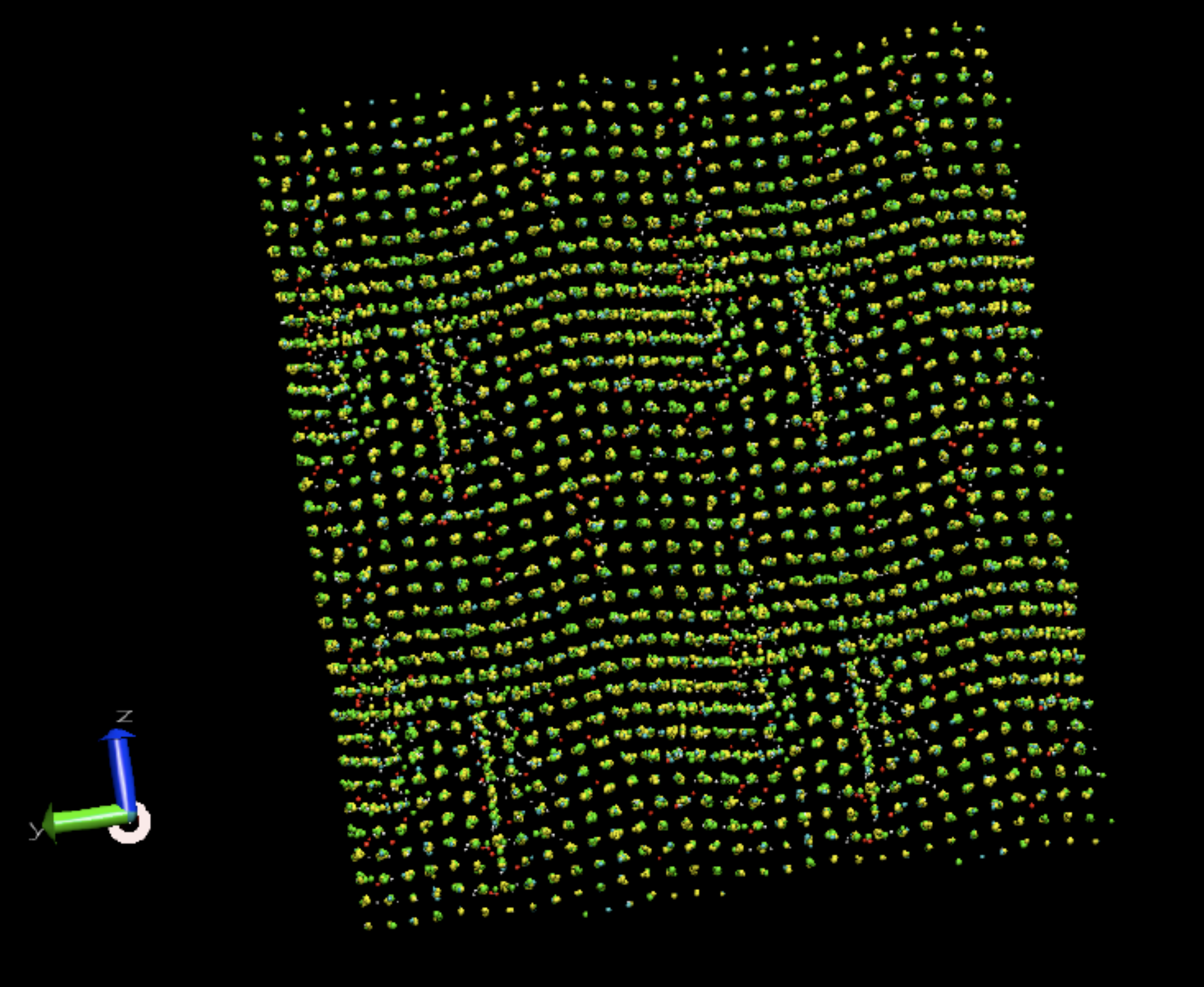}
\caption{(Color on line) Initial Configuration of the 27648 ion mixture as described in the text.}
\label{Fig1}
\end{center}
\end{figure}

We start from initial conditions where we try and minimize arbitrary assumptions about the distribution of impurities (low $Z$ ions) in the solid.  A small 432 ion system, with composition from Table \ref{tableone}, is started from random positions at a high temperature and a relatively high reference density $n=7.18\times 10^{-5}$ fm$^{-3}$.  Results can be scaled to other densities at constant $\Gamma$, Eq. \ref{gamma}.  The system is then cooled until it is observed to freeze.  Note that it is straight forward to crystalize such a small system.  Next eight copies of the 432 ion solid are assembled into a 3456 ion configuration and this is evolved for a short time.   Finally eight copies of this 3456 ion system are assembled into the final 27648 ion system.  The configuration of this system is shown in Fig. \ref{Fig1}.  The wavy planes of ions in Fig. \ref{Fig1} show that the crystal is strained and it is not in equilibrium.  

\begin{figure}[ht]
\begin{center}
\includegraphics[width=3in,angle=0,clip=true] {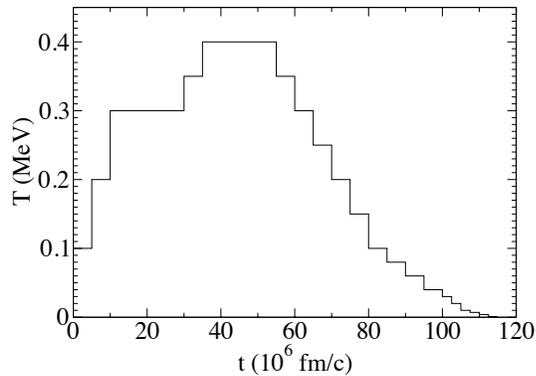}
\caption{Cooling schedule of temperature $T$ versus time $t$ for an MD simulation of a 27648 ion system, see text.}
\label{Fig2}
\end{center}
\end{figure}

\begin{figure}[ht]
\begin{center}
\includegraphics[width=3.2in,angle=0,clip=true] {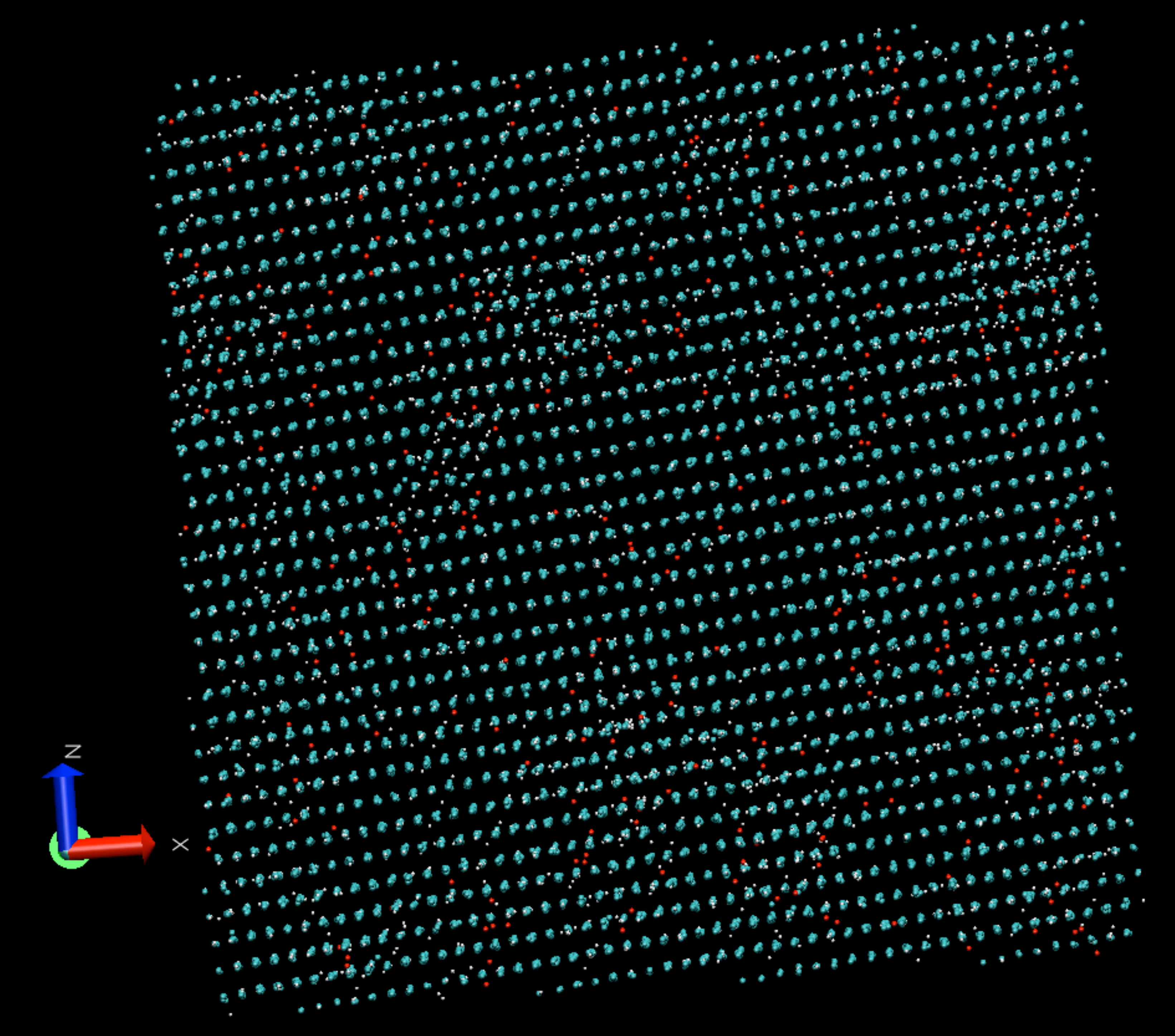}
\caption{(Color on line) Final configuration of the 27648 ion system at near zero temperature.  The medium sized red spheres are Oxygen ions, while the small white spheres are other below average $Z$ ions, and the large blue spheres are above average $Z$ ions.}
\label{Fig3}
\end{center}
\end{figure}

\begin{figure}[ht]
\begin{center}
\includegraphics[width=3.3in,angle=0,clip=true]{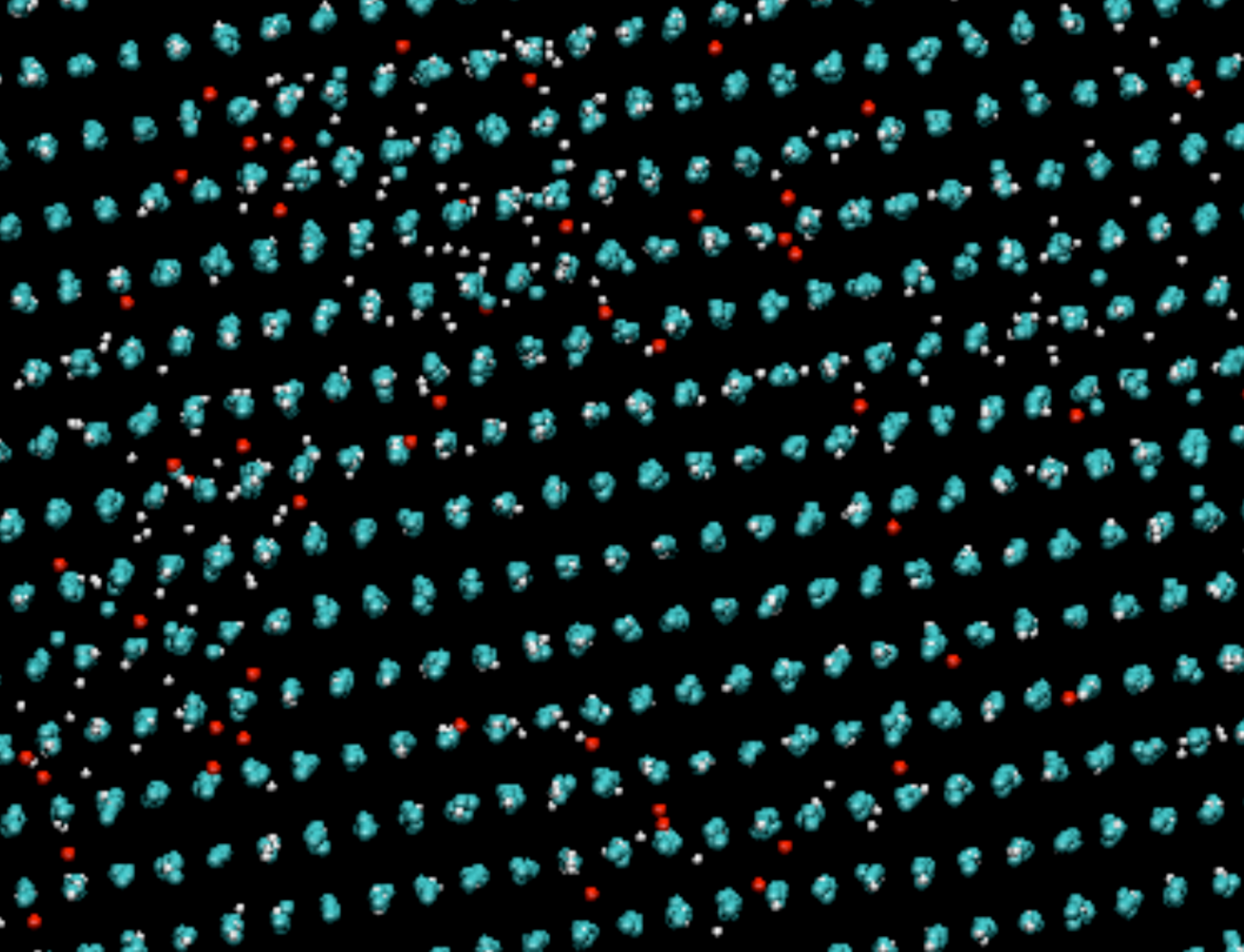}
\caption{(Color on line) An enlarged detail from Fig. \ref{Fig3}.}
\label{Fig4}
\end{center}
\end{figure}

\begin{figure}[ht]
\begin{center}
\includegraphics[width=3.6in,angle=0,clip=true] {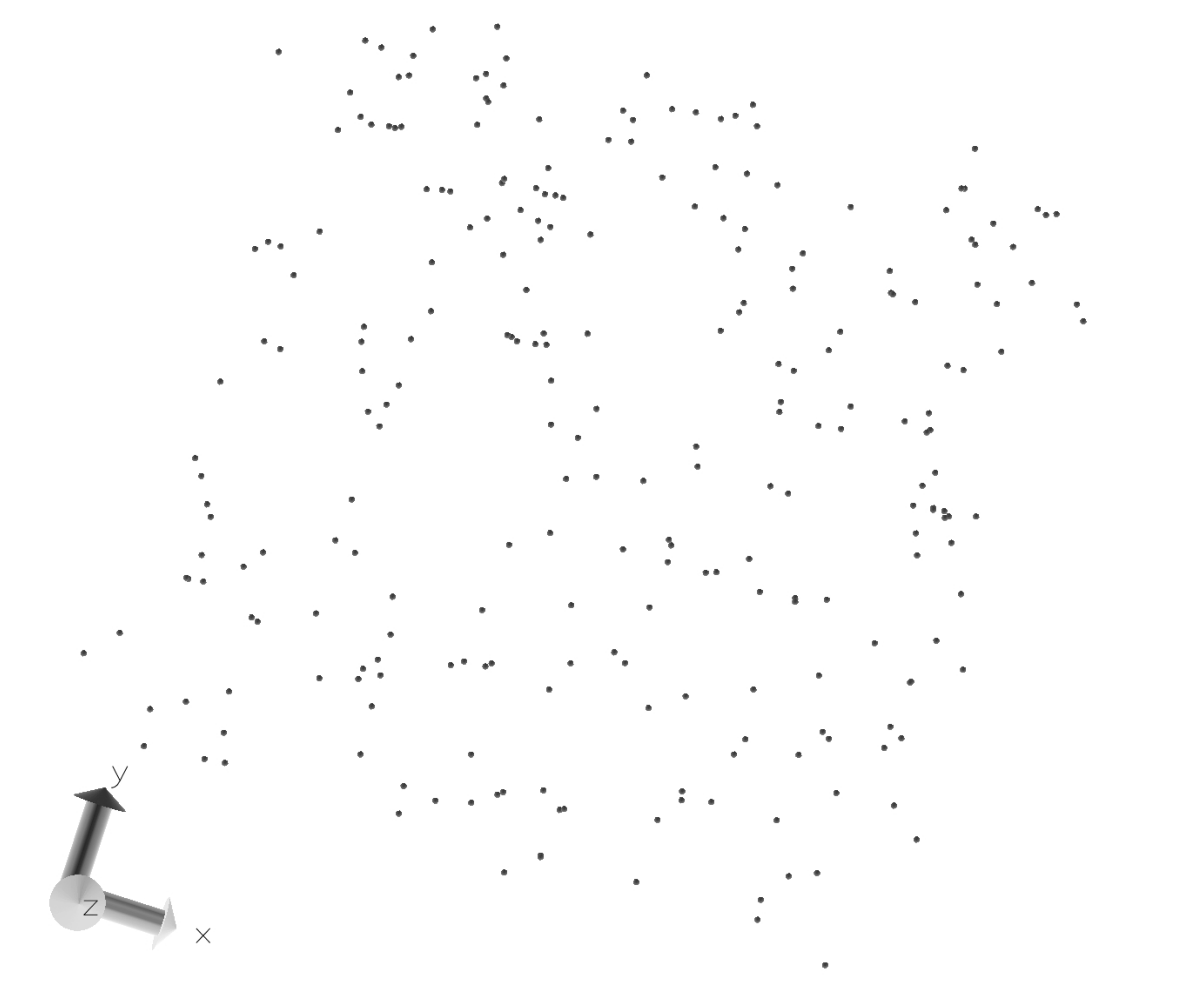}
\caption{Final configuration of only the 256 Oxygen ions (out of the total of 27648 ions ) in the system at near zero temperature. }
\label{Fig5}
\end{center}
\end{figure}

We anneal this 27648 ion system by evolving it according to the cooling schedule in Fig. \ref{Fig2}.  The simulations were performed on a special purpose MDGRAPE-2 board \cite{mdgrape} provided by Indiana University's High Performance Computing group and took 46 days.  The system is first heated to near the melting point and then cooled to near zero temperature.  The final configuration is shown in Figs. \ref{Fig3} and \ref{Fig4}.   We see that the system forms a regular crystal with a large distribution of impurities.  For example, Fig. \ref{Fig5} shows the configuration of only the Oxygen ions.  These are seen to be distributed throughout the simulation volume.  However, there are strong correlations between the ions that will be discussed in the next section.

\section{Results}
\label{Results}

\begin{figure}[ht]
\begin{center}
\includegraphics[width=3.75in,angle=0,clip=true] {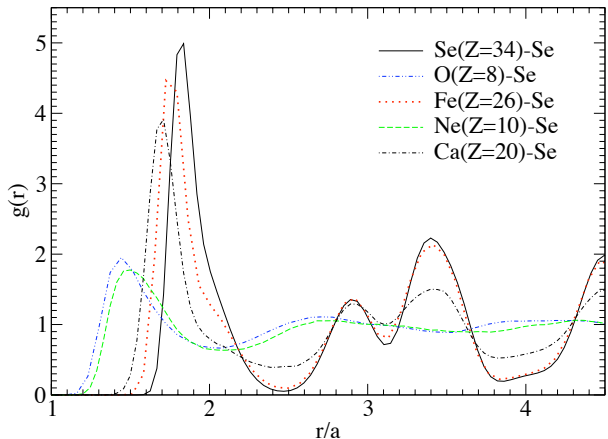}
\caption{(Color on line) Radial distribution functions $g_{ij}(r)$ for ions of type $i$=Se and type $j$  versus r over the mean ion sphere radius $a$ at a temperature $T=0.1$ MeV.  The solid line shows $g(r)$ for Se-Se correlations while the other curves show correlations between Se and O (dash-dot-dot), Ne (dashed), Ca (dot-dashed), and Fe (dotted).}
\label{Fig6}
\end{center}
\end{figure}

We now present results for the radial distribution function and static structure factor to characterize the distribution of impurities in the sample.  The radial distribution function $g_{ij}(r)$ is the probability of finding an ion of type $j$ a distance $r$ away from a given ion of type $i$.  It is normalized to go to one for large $r$.    We calculate $g_{ij}$ by histogramming relative distances for 2500 MD configurations of the 27648 ions where each configuration is separated by a time of 250 fm/c.  The dominant species is Se ($Z=34$), see Table \ref{tableone}.   Figure \ref{Fig6}  shows the diagonal $g_{ii}(r)$ for Se-Se correlations.  This shows peaks corresponding to the dominant body centered cubic lattice structure.  In addition there are dips near $r/a=2.5$ and 4 with $a$ the mean ion sphere radius.  Figure \ref{Fig6} also shows $g_{ij}(r)$ for correlations between Fe and Se ions.  This is very similar to that for Se-Se.  We conclude that most Fe impurities are substitutional and occupy vacant Se lattice sites.  In contrast, Fig. \ref{Fig6} shows O-Se and Ne-Se correlation functions are significantly different and do not show dips near $r/a=2.5$ and 4.  This suggests that most O and Ne impurities are interstitial and occupy positions between occupied Se lattice sites.  This can be understood if low $Z$ impurities are ``smaller'' than Se ions because reduced Coulomb repulsion allows them to fit into interstitial positions.  Finally Ca is seen to be an intermediate case.  Calcium impurities may occupy both substitutional and interstitial sites.

\begin{figure}[ht]
\begin{center}
\includegraphics[width=3.75in,angle=0,clip=true] {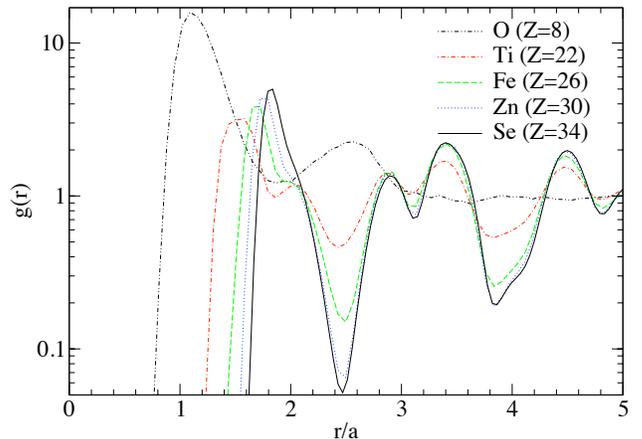}
\caption{(Color on line) Radial distribution functions $g_{ii}(r)$ for diagonal correlations between ions of type $i$ versus r over the mean ion sphere radius $a$ at a temperature $T=0.1$ MeV.  The solid line shows $g(r)$ for Se-Se correlations while the other curves show correlations between O-O (dash-dot-dot), Ti-Ti (dash-dot), Fe-Fe (dashed), and Zn-Zn (dotted).}
\label{Fig7}
\end{center}
\end{figure}

\begin{figure}[ht]
\begin{center}
\includegraphics[width=3.75in,angle=0,clip=true] {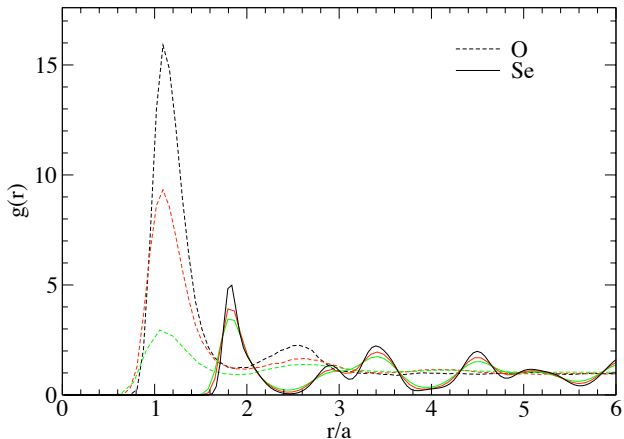}
\caption{(Color on line) The temperature dependence of the radial distribution functions $g_{ii}(r)$ versus r over the mean ion sphere radius $a$ at temperatures of (bottom to top) $T=0.3 (green) $, 0.2 (red) , and 0.1 (black) MeV.  The solid lines show $g(r)$ for Se-Se correlations while dashed curves show  O-O correlations.}
\label{Fig8}
\end{center}
\end{figure}

Figure \ref{Fig7} shows diagonal $g_{ii}(r)$ for O-O, Ti-Ti, Fe-Fe, and Zn-Zn correlations as well as Se-Se correlations as in Fig. \ref{Fig6}.  The O-O correlation is seen to have a very large peak near $r/a=1$, note the log scale.  This peak is consistent with the clustering visible in Fig. \ref{Fig5}.  This suggests that a number of ``small'' O ions may cluster together and take the place of a larger $Z$ ion.  To study this peak further we show its temperature dependence in Fig. \ref{Fig8}.   The peaks in the Se-Se correlation function grow sharper with decreasing temperature as the amplitude of oscillations is reduced.  In contrast the area under the first peak in the O-O correlation function grows rapidly as the temperature decreases.  This shows that the O ions are becoming very strongly correlated at lower temperatures.   This could greatly increase the rate of some pycnonuclear reactions \cite{yakovlev}\cite{pycno}.   Note that because of these strong correlations, it is possible that the O impurities may not have fully equilibrated during our simulation.  Indeed it is even possible that they could phase separate at low temperatures, although this could take a very long simulation time.

The static structure factor $S(q)$ describes electron-ion scattering.  This is important for transport properties such as the thermal conductivity, electrical conductivity, or shear viscosity.  We calculate $S(q)$ directly as a density-density correlation function using trajectories from our MD simulations,
\begin{equation}
S({\bf q})=\langle \rho^*({\bf q})\rho({\bf q}) \rangle - |\langle \rho({\bf q})\rangle |^2\, .
\label{S(q)}
\end{equation}
Here the charge density $\rho({\bf q})$ is,
\begin{equation}
\rho({\bf q}) = \frac{1}{\sqrt N} \sum_{i=1}^N \frac{Z_i}{\langle Z \rangle} {\rm e}^{i {\bf q \cdot r}_i},
\label{rho}
\end{equation}
with $N$ the number of ions in the simulation and $Z_i$, ${\bf r}_i$ are the charge and location of the ith ion.  We evaluate the thermal average in Eq. \ref{S(q)} as a time average during our MD simulations.  We use 2500 configurations that are separated in time by 250 fm/c.  We also average over the direction of the vector ${\bf q}$.  This calculation is somewhat time consuming because of the very large number of separate ${\bf q}$ values involved.    

\begin{figure}[ht]
\begin{center}
\includegraphics[width=3.75in,angle=0,clip=true] {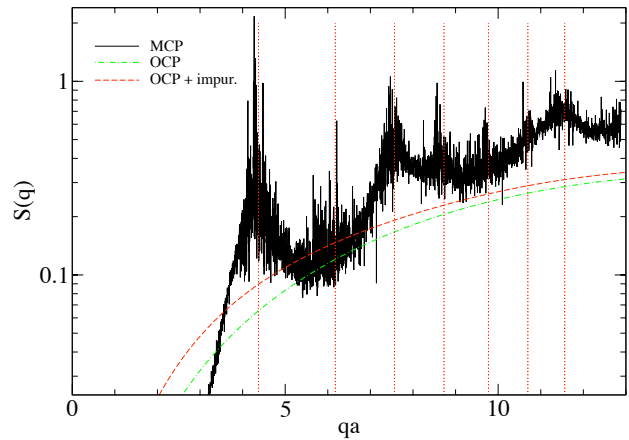}
\caption{(Color on line) The static structure factor $S(q)$ versus momentum transfer $q$ times mean ion sphere radius $a$.  Our MD simulation is the black curve.  A simple fitting formula for a one component plasma is the green dash-dotted line while the red dashed line includes impurity scattering from uncorrelated impurities.  The vertical red dotted lines show the positions of Bragg peaks for scattering from a pure one component plasma body centered cubic lattice. }
\label{Fig9}
\end{center}
\end{figure}

In Fig. \ref{Fig9} we show $S(q)$ at a temperature of 0.1 MeV.  Although our result has some statistical noise, we see peaks that correspond to Bragg scattering from the crystal lattice.  We also show in Fig. \ref{Fig9}, as vertical lines of arbitrary height, the positions of Bragg peaks expected for a body centered cubic lattice of a one component system.  The pattern of peaks for our $S(q)$ confirms that our lattice is also body centered cubic.  However, the first peak near $qa=4.2$ occurs at a slightly smaller $q$ than that for a OCP.  This indicates that our unit cell is slightly larger and contains slightly more than two ions.  This is because some of the low $Z$ impurities occupy interstitial lattice sites.  Also shown is a simple fit to one component plasma (OCP) results from ref. \cite{OCPfit}.  Note that the OCP fit averages over sharp structures, see for example ref. \cite{ocp_calc}.  In addition Fig. \ref{Fig9} shows $S(q)$ for a OCP plus the contribution of impurity scattering assuming the impurities are almost uncorrelated as in ref. \cite{impurities}.  This curve assumes an impurity parameter $Q=22.54$ and is below our full $S(q)$ result.  This suggests that the important correlations that we find between impurities, see for example Fig. \ref{Fig8}, increase the contribution of impurity scattering to $S(q)$.

From our $S(q)$ results we calculate the thermal conductivity $\kappa$ as in ref. \cite{thermalcond}.  
\begin{equation}
\kappa=\frac{\pi k_B^2T k_F}{12 \alpha^2 \langle Z \rangle \Lambda}
\end{equation}
Here $k_B$ is the Boltzmann constant, $k_F$ the electron Fermi momentum, $\alpha $ the fine structure constant and the coulomb logarithm $\Lambda$ is,
\begin{equation}
\Lambda=\int_{q_0}^{2k_F}\frac{dq}{q\epsilon(q,0)^2}S^\prime(q)(1-\frac{q^2}{4k_F^2}).
\label{Lambda}
\end{equation}
Here $\epsilon$ is the dielectric function of the electrons and $S^\prime(q)$ is the inelastic part of $S(q)$, see ref. \cite{thermalcond}.   Our results for $\Lambda$ and $\kappa$ are listed in Table \ref{tabletwo} where we also show results ($\Lambda_{\rm imp}$, $\kappa_{\rm imp}$) for the simple fit to OCP results plus scattering from uncorrelated impurities.  We find that $\kappa$ is somewhat reduced compared to $\kappa_{\rm imp}$ because of the increased contributions of impurity scattering.

\begin{table}
\caption{Thermal conductivity $\kappa$ results for a temperature of $T=0.043$ MeV ($5\times 10^8$K). The original runs were at a reference density of $10^{13}$ g/cm$^3$ and the indicated temperatures $T$.    The Coulomb logarithm $\Lambda$ is defined in Eq. \ref{Lambda}. These results have been scaled to the indicated densities.} 
\begin{tabular}{lllllll}
$T$ & $\Gamma$ & $\Lambda$ & $\Lambda_{\rm imp}$   & $\rho$& $\kappa$ & $\kappa_{\rm imp}$ \\
\toprule
MeV &  & & & g/cm$^3$ & erg/K cm s & erg/K cm s \\
0.1 & 910 & 0.160 & 0.121 & $8.0\times 10^{11}$ & $1.57\times 10^{19}$ & $2.08\times 10^{19}$\\
0.2 & 455 & 0.299 & 0.253 & $9.9\times 10^{10}$ & $4.20\times 10^{18}$ & $4.96\times 10^{18}$ \\
0.3 & 304 & 0.418 & 0.345 & $2.9\times 10^{10}$ & $2.00\times 10^{18}$ & $2.42 \times 10^{18}$\\
\end{tabular} 
\label{tabletwo}
\end{table}

\section{Summary and Conclusions}
\label{Conclusions}
Using molecular dynamics simulations we have calculated the structure of a 27648 ion crystal made from a complex composition of rapid proton capture nucleosynthesis ash that includes many impurities.  Even with many impurities characterized by a large impurity parameter $Q=22.54$, we find a regular body centered cubic crystal with long range order.  We do not find an amorphous structure.  We find that low $Z$ impurities often occupy interstitial sites while high $Z$ impurities tend to occupy substitutional sites.  There are strong attractive short range correlations between low $Z$ impurities that grow with decreasing temperature.  These correlations could significantly enhance the rate of pycnonuclear (density driven) reactions at high densities.

The static structure factor $S(q)$ is enhanced by impurity scattering over $S(q)$ for a one component plasma.  Furthermore there are important correlations between impurities that may invalidate simple models that assume the impurities are randomly distributed.  The thermal conductivity is reduced by impurity scattering to such an extent that if the inner crust is as impure as the present simulations for the outer crust show, then the electron thermal conductivity in the inner crust will be reduced to such an extent that it may disagree with interpretations \cite{rutledge,shternin,cumming08} of observations \cite{cackett} of rapid crust cooling.   If the interpretation of these observations is correct, we conclude that either (a) the large impurity concentrations in our initial rp ash composition are wrong or (b) impurity concentrations are reduced by the time material is buried deeper into the inner crust.  This could be because of nuclear reactions.

Finally these results will be used in additional work to calculate the impact of impurities on the mechanical properties of the crust including the shear modulus and the breaking strain.  The shear modulus is important for crust oscillations and the breaking strain may be important for crust breaking models of magnetar giant flares and for the stability of mountains that may radiate gravitational waves from rapidly rotating neutron stars.

\section{Acknowledgments}
This work was supported in part by DOE grant DE-FG02-87ER40365 and by Shared University Research grants from IBM, Inc. to Indiana University.

\vfill\eject


\begin{thebibliography}{99} 
\bibitem{yukawasystems} S. Hamaguchi, R. T. Farouki, D. H. E. Dubin, Phys. Rev. E {\bf 56}, 4671 (1997).

\bibitem{rp1} H. Schatz et al., PRL {\bf 86} (2001) 3471.

\bibitem{rp2} S. E. Woosley, A. Hager, A. Cumming, R. D. Hoffman, J. Pruet, T. Rauscher, J. L. Fisker, H. Schatz, B. A. Brown, and M. Wiescher, ApJ Supp. {\bf 151} (2004) 75.
\bibitem{gupta} S. Gupta, E. F. Brown, H. Schatz, P. Moller, and K-L. Kratz, ApJ {\bf 662} (2007) 1188.
\bibitem{mcocp} H. E. Dewitt, W. L. Slattery, and J. Yang in ``Strongly Coupled Plasmas'', eds. H. M. Van Horn and S. Ichimaru, Univ. of Rochester Press 1993, p425.
\bibitem{MCPliquid} K. Wunsch, P. Hilse, M. Schlanges, D. O. Gericke, Phys. Rev. E {\bf 77}, 056404 (2008).
\bibitem{yakovlev} D. G. Yakovlev, L. R. Gasques, M. Beard, M. Wiescher, and A. V. Afanasjev, PRC {\bf 74} (2006) 035803.
\bibitem{LJglass} W. Kob and J.-L. Barrat, Phys. Rev. Lett. {\bf 78}, 4581 (1997).

\bibitem{horowitz} C. J. Horowitz, D. K. Berry, and E. F. Brown, Phys. Rev. E {\bf 75}, 066101 (2007).

\bibitem{pycno} C. J. Horowitz, H. Dussan and D. K. Berry, Phys. Rev. C {\bf 77}, 045807 (2008). 

\bibitem{thermalcond} C. J. Horowitz, O. L. Caballero, and D. K. Berry, Phys. Rev. E {\bf 79}, 026103 (2009).

\bibitem{Wijnands} R.~Wijnands et al., astro-ph/0405089.

\bibitem{cackett} E. M. Cackett et al., MNRAS {\bf 372}, 479 (2006).

\bibitem{rutledge} R. E. Rutledge et al., ApJ. {\bf 580}, 413 (2002).

\bibitem{shternin} P. S. Shternin, D. G. Yakovlev, P. Haensel, and A. Y. Potekhin, MNRAS {\bf 382}, L43 (2007).
\bibitem{cumming08} Edward F. Brown and Andrew Cumming, arXiv:0901.3115.
\bibitem{shearmod} C. J. Horowitz and J. Hughto, arXiv:0812.2650.
\bibitem{ogata} Shuji Ogata and Setsuo Ichimaru, Phys. Rev. A {\bf 42} (1990) 4867.
\bibitem{qpos} Lars Samuelsson, Nils Andersson, {\it Mon. Not. Roy. Astron. Soc.} {\bf 374}, 256 (2007). 
\bibitem{gw} G. Ushomirsky, C. Cutler, and L. Bildsten, MNRAS {\bf 319} (2000) 902.  A. L. Watts, B. Krishnan, L. Bildsten, and B. F. Schutz, MNRAS {\bf 389} (2008) 839.
\bibitem{thompsonduncan}C. Thompson, R. C. Duncan, {\it Astrophys. J.} {\bf 561}, 980 (2001).
\bibitem{kadau}  C. J. Horowitz and K. Kadau, Phys. Rev. Let. {\bf 102}, 191102 (2009).
\bibitem{impurities} N. Itoh and Y. Kohyama, ApJ. {\bf 404} (1993) 268.
\bibitem{mdgrape} T. Narumi, R. Susukita, T. Ebisuzaki, G. McNiven, and B. Elmegreen, Molecular Simulation 21, 401 (1999).
\bibitem{OCPfit} A. Y. Potekhin, D. A. Baiko, P. Haensel, and D. G. Yakovlev, Astron. Astrophysics {\bf 346}, 345 (1999).
\bibitem{ocp_calc} D. A. Baiko, D. G. Yakovlev, H. E. DeWitt, and W. L. Slattery, Phys. Rev. E {\bf 61}, 1912 (2000).




\end{thebibliography}
\end{document}